\documentclass[review]{elsarticle}
\usepackage{natbib}
\usepackage{graphicx}
\usepackage{float}
\usepackage{gensymb}
\usepackage[mathscr]{euscript}
\usepackage{amsmath}
\usepackage{siunitx}
\usepackage[dvipsnames]{xcolor}

\begin{document}

\date{\today}

\begin{frontmatter}

\title{Phase Slips and Metastability in Granular Boron-doped Nanocrystalline Diamond Microbridges}

\author[cdf]{G.M. Klemencic \corref{cor1}} \ead{KlemencicG@cardiff.ac.uk}
\address[cdf]{
    School of Physics and Astronomy, 
    Cardiff University, 
    Queen's Buildings, 
    The Parade, Cardiff, CF24 3AA, UK
    }
\author[brm]{D. T. S. Perkins}
\address[brm]{
    School of Physics and Astronomy, 
    University of Birmingham, Birmingham, B15 2TT, UK
    }
\author[bst]{J. M. Fellows}
\address[bst]{
    School of Physics, HH Wills Physics Laboratory, 
    University of Bristol, Tyndall Avenue, 
    Bristol, BS8 1TL, UK
    }

\author[brm]{C. M. Muirhead}
\author[brm]{R. A. Smith}
\author[cdf]{S. Mandal}
\author[cdf]{S. Manifold}
\author[cdf]{M. Salman}
\author[cdf]{S. R. Giblin}
\author[cdf]{O. A. Williams}

\cortext[cor1]{Corresponding author}

\begin{abstract}
A phase slip is a localized disturbance in the coherence of a superconductor allowing an abrupt 2$\pi$ phase shift. Phase slips are a ubiquitous feature of one-dimensional superconductors and also have an analogue in two-dimensions. Here we present electrical transport measurements on boron-doped nanocrystalline diamond (BNCD) microbridges where, despite their three-dimensional macroscopic geometry, we find clear evidence of phase slippage in both the resistance-temperature and voltage-current characteristics. We attribute this behavior to the unusual microstructure of BNCD. We argue that the columnar crystal structure of BNCD forms an intrinsic Josephson junction array that supports a line of phase slippage across the microbridge. The voltage-state in these bridges is metastable and we demonstrate the ability to switch deterministically between different superconducting states by applying electromagnetic noise pulses. This metastability is remarkably similar to that observed in $\delta$-MoN nanowires, but with a vastly greater response voltage. 
\end{abstract}

\end{frontmatter}


\section{Introduction}{\label{Introduction}}

Superconductivity in reduced dimensions is the subject of resurgent interest for both fundamental and practical reasons. Historically, phase slips were considered an unwelcome experimental feature; however recent work on coherent quantum phase slips\cite{astafiev2012coherent} has revealed promising device applications that exploit the duality between these and the Josephson effect\cite{de2018charge} in both superconducting quantum computing\cite{mooij2006superconducting} and metrology\cite{wang2019towards}. The length scale governing electronic dimensionality in superconductors is the coherence length, $\xi(T)$, which describes the shortest length over which the superconducting wavefunction may vary. If a sample thickness or width is reduced below $\xi(T)$, the electronic dimensionality is correspondingly reduced. Two-dimensional superconductivity has been observed in ultrathin films\cite{haviland1989onset,lin2015superconductivity}, atomically thin exfoliated single crystals\cite{paradiso2019phase,saito2017highly}, and heterogeneous interfaces \cite{reyren2007superconducting}. One-dimensional superconductivity has been seen in nanowires and whiskers\cite{arutyunov2008superconductivity,bezryadin2000quantum}, whilst zero-dimensional superconductivity has been seen in granular samples of superconducting grains in a non-superconducting matrix \cite{klemencic2017fluctuation,li2003quantum}. 

Superconductivity in reduced dimensions is of particular interest for the diamond community as this material has notably been shown to display the properties of 0-, 2-, and 3-dimensional superconductivity in different experiments, as described below. In this paper, we will show behavior reminiscent of 1D superconductivity in samples far from that limit. 
We will argue that the origin of this behavior is a macroscopic analog of the phase slips in truly 1D systems brought about by nanocrystalline diamond's unusual microstructure, making diamond a promising candidate for novel devices exploiting the applications  of phase slips.

A distinctive feature of 1D superconductivity is a residual resistance below the transition temperature, $T_c$ \cite{langer1967intrinsic,mccumber1970time}. In the low-current ohmic limit, this is caused by phase slips\cite{skocpol1974phase}. For a phase slip to occur, a free energy barrier must be overcome either thermally, or tunneled through quantum mechanically, leading to thermally activated and quantum phase slips, respectively. In 1D nanowires, suppression of the order parameter interrupts superconducting transport, but Cooper pairs can tunnel across the phase slip leading to a finite conductance in the superconducting state\cite{bezryadin2000quantum}. At intermediate currents, non-equilibrium phase slip centers (PSC) form\cite{skocpol1974phase}, whose signature is the appearance of discrete voltage steps in the voltage-current $V(I)$ characteristic above some onset current, due to successive PSC formation along the sample length\cite{li2015local,delacour2012quantum}.

\begin{figure}
\includegraphics[width=0.8\textwidth]{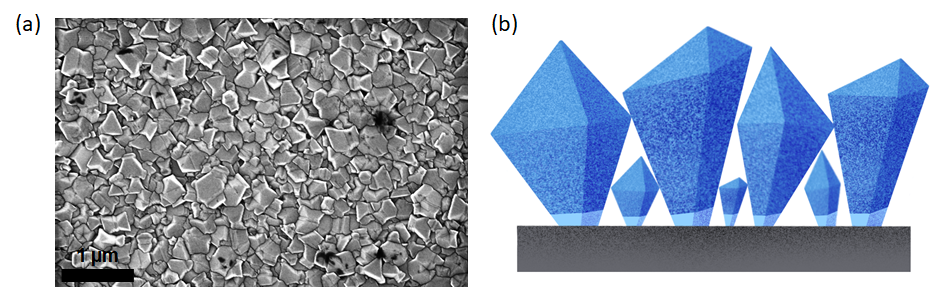}
\caption{\textbf{Microstructure of boron doped nanocrystalline diamond films grown by microwave plasma-assisted chemical vapor deposition.}
\textbf{(a)} Scanning electron micrograph of a 339\,nm thick BNCD film surface with clearly defined grains with an average size of 102\,nm.
\textbf{(b)} Schematic of the microstructure of the film showing columnar grain growth originating from the (light blue) seed crystals.}
\label{sem}
\end{figure}

Here we present low temperature electronic transport measurements of photolithographically defined 3D boron-doped nanocrystalline diamond (BNCD) microbridges that bear the signatures of phase slip physics. BNCD films can be grown by chemical vapor deposition on non-diamond substrates by seeding the substrate with nanodiamond particles \cite{williams2007enhanced}. These seed crystals grow epitaxially, growing laterally and vertically to form 3D islands until a fully coalesced film forms~\cite{jiang1994nucleation}. A competitive columnar growth process then proceeds following the Van der Drift model\cite{smereka2005simulation}. The resulting film comprises columnar superconducting diamond crystals separated by non-superconducting grain boundaries\cite{dahlem2010spatially}. A scanning electron micrograph is shown alongside a schematic columnar growth structure in Fig.~\ref{sem}. In a previous publication\cite{klemencic2017fluctuation}, we have shown a cross-sectional SEM image of a BNCD film grown under the same conditions as the sample used here. Fig.~1b of that paper shows a grain boundary spanning the entire vertical extent of the film. Other works\cite{dahlem2010spatially, zhang2019anomalous} have also confirmed the columnar nature of the grains in similarly grown films. The general superconducting properties of BNCD ($T_c \sim 4$\,K and $\xi\sim10$\,nm)\cite{klemencic2017fluctuation} are similar to bulk single crystalline samples\cite{bustarret2004dependence} but the detailed behavior is modified by the granular microstructure\cite{dahlem2010spatially,zhang2019anomalous}, which is unusual for a superconducting material in that columnar grains extend vertically through the entire film. Indeed, we have previously seen a 3D--0D--3D dimensional crossover in the fluctuation conductivity arising from the granularity\cite{klemencic2017fluctuation,klemencic2019observation}.

To understand how these 3D microbridges can apparently show 1D behavior, we note that PSCs have higher dimensional analogues. Their 2D analog, phase slip lines (PSLs), have been observed in wide superconducting strips \cite{sivakov2003josephson,dmitriev2005critical,bell2007one,paradiso2019phase}. Although their experimental signatures are similar to PSCs, their physical mechanism is different since it is too energetically costly to suppress the order parameter across the whole width, $w\gg\xi(T)$, of the strip. Numerical simulations based on time-dependent Ginzburg-Landau theory suggest that PSLs can be caused by fast-moving vortices traveling perpendicular to the current flow\cite{berdiyorov2014dynamics}. These simulations, confirmed experimentally\cite{sivakov2003josephson}, predict that PSLs behave as dynamically created Josephson junctions\cite{paradiso2019phase}. Like PSCs, PSLs result in a resistance below $T_c$\cite{bell2007one} and discrete voltage steps in the $V(I)$ characteristic\cite{dmitriev2005critical,paradiso2019phase}.

Remarkably, we observe the hallmarks of phase slip phenomena described above in BNCD microbridges, despite their width and thickness being well within the 3D limit. Given the morphology, with columnar superconducting grains in a normal matrix, we may expect to see transport phenomena associated with a randomly disordered 2D Josephson junction array\cite{yu1992resistance,tighe1991vortex}. This is supported by behavior reminiscent of a BKT transition -- a distinctly 2D phenomenon -- observed in similar BNCD films\cite{coleman2017possible}. We conclude that the signatures of low-dimensional superconductivity must therefore arise from the microstructure of the material. We present, first our results, and then return to this model to consider the extent to which phase slip theory provides an explanation.


\section{Results}

\subsection{Resistance in the superconducting state}

\begin{figure}
\includegraphics[width=\textwidth]{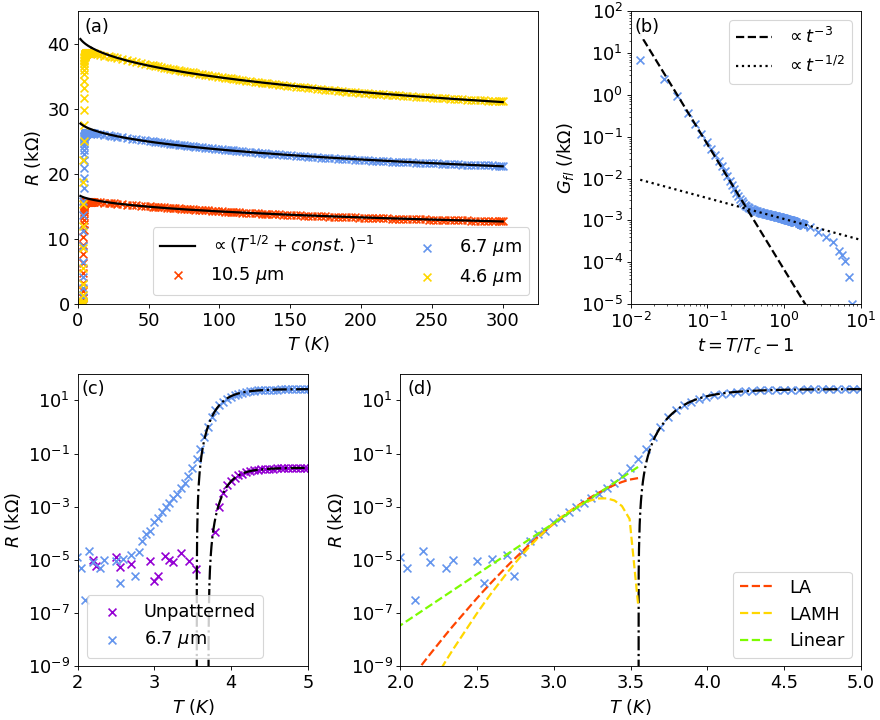}
\caption{\textbf{Resistance as a function of temperature, $R(T)$, for BNCD microbridges - focusing on the 6.7\,$\mu$m wide microbridge -  and the unpatterned film.}
\textbf{(a)} $R(T)$ for three 1600\,$\mu$m long microbridges (widths as indicated by the legend) in the temperature region 2 -- 300\,K. 
The high temperature data are fit to the form $G(T)=a+b\sqrt{T}$ expected from electron-electron interaction theory. 
\textbf{(b)} Log-log plot of the fluctuation conductivity $G_{fl}(T)$ as a function of the reduced temperature, showing a crossover from 0D to 3D behavior at $T-T_c\sim 0.3$\,K.
\textbf{(c)} The superconducting transition of the unpatterned film does not show evidence of broadening below $T_c$. The black line is a fit to the fluctuation conductivity. The superconducting transition of the 6.7\,$\mu$m wide microbridge is shown for comparison.
\textbf{(d)} Low temperature $R(T)$ for the 6.7\,$\mu$m wide microbridge. The resistance below $T_c$ is fit to the LA (red), LAMH (yellow) and
linear (green) forms. Note that none of these are expected to be accurate close to $T_c$ as they assume large free energy barriers.}
\label{RvsT}
\end{figure}

The resistance as a function of temperature, $R(T)$, of three BNCD microbridges of different widths is shown in Fig.~\ref{RvsT}(a). The microbridges are fabricated from the same BNCD film which has a thickness of 339\,nm and an average grain size of 102\,nm. Each microbridge has a length, $L$, of 1600\,$\mu$m and their widths, $w$, are 4.6, 6.7, and 10.5\,$\mu$m. The geometric dimension of each of these microbridges is many times larger than the typical coherence length reported for BNCD ($\sim$10\,nm) and therefore they are considered to be three-dimensional structures. 

At temperatures up to the measurement limit, 300\,K, the normal state conductance has the form $G(T)=a+b\sqrt{T}$, as expected from granular electron-electron interaction (EEI) theory\cite{beloborodov2007granular} in three-dimensions. This term is subtracted from the measured conductance over the whole temperature range to leave only the fluctuation conductance, $G_{fl}(T)$, near $T_c$. When $G_{fl}(T)$ is plotted against the reduced temperature, $t=(T-T_c)/T_c$, on a log-log plot (Fig.~\ref{RvsT}b), three distinct regions are expected\cite{klemencic2017fluctuation}, with power laws $-\tfrac{1}{2}$, $-3$, and $-\tfrac{1}{2}$, respectively as $T_c$ is approached. Note that
in Fig.~\ref{RvsT}(b) only the transition furthest from $T_c$ is distinctly seen; the power law behavior is only expected close to $T_c$, which is why the
results deviate from the $-\tfrac{1}{2}$ power law at higher temperature. The $(T-T_c)^{-3}$ region is the widest, and fitting to this allows very accurate
determination of $T_c$. A residual resistance is seen
below $T_c$ for the three microbridges (Fig.~\ref{RvsT}d and Supplementary Information), but not for the unpatterned film (Fig.~\ref{RvsT}c), strongly suggesting that some form of phase slip phenomenon is taking place in these patterned microbridges\cite{bezryadin2000quantum,mikheenko2005phase}.

Despite the fact that these films are clearly three-dimensional, as evidenced by the fitting to the EEI theory, the residual resistance below $T_c$ has features reminiscent of the Langer-Ambegaokar-McCumber-Halperin (LAMH)\cite{langer1967intrinsic, mccumber1970time} behaviors commonly seen in one-dimensional superconductors. The appearance of 1D phenomena in higher dimensional systems has recently been reported in systems such as percolating films of Pb nanoparticles\cite{nande2017quantum}, and Pb$_x$(SiO$_2$)$_{1-x}$ nanocrystalline 
films\cite{duan2019hopping}. The various theories of thermally activated phase slips all predict a resistance below $T_c$ of the activated tunneling form

\begin{equation}
R(T)=\frac{\hbar^2}{e^2}\frac{\Omega(T)}{k_B T}\exp\left(-\frac{\Delta F(T)}{k_B T}\right),
\label{RT_slip}
\end{equation}


\noindent where $\Omega(T)$ is an attempt rate, and $\Delta F(T)$ is a free energy activation barrier. In the region not too close to $T_c$, $\Omega(T)=a(T_c-T)^\alpha$ and $\Delta F(T)=b(T_c-T)^\beta$, where the powers $\alpha$, $\beta$, are theory dependent and $a$ and $b$ are fitting parameters. In Fig.~\ref{RvsT}d, we fit the $R(T)$ below $T_c$ for the 6.7\,$\mu$m wide bridge to the Langer-Ambegaokar (LA), LAMH, and linear forms for which $(\alpha,\beta)$ are $(0,\tfrac{3}{2})$, $(\tfrac{9}{4},\tfrac{3}{2})$ and $(0,1)$, respectively. The data in the resistance range from 10$^{-5}\,k\Omega$ to 10$^{-2}\,k\Omega$ appears to fit best to the linear form. Below $10^{-5}\,k\Omega$ we are limited by noise, whilst above 10$^{-2}\,k\Omega$ we are too close to $T_c$ for these theories to work, and some numerical interpolation formula is needed. The linear form is equivalent to the inverse Arrhenius form reported in granular Pb, Sn and Pb-Ag films\cite{frydman2002universal}.  A resistance below $T_c$, whose temperature dependence shows thermally activated tunneling, is indicative of the presence of phase slips in these microbridges.


\subsection{Voltage-current characteristics}

\begin{figure}
\includegraphics[width=\textwidth]{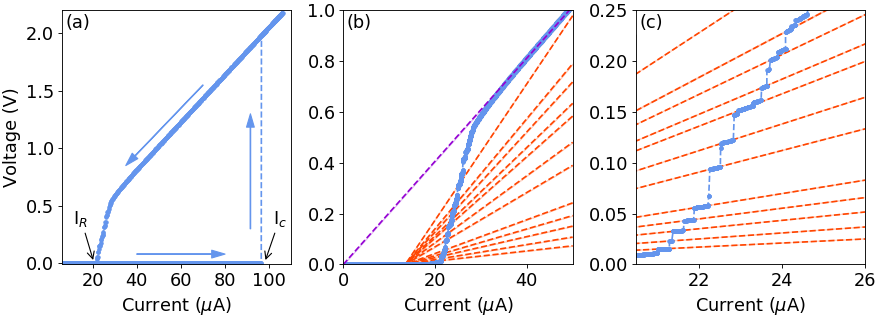}
\caption{
\label{IVC}
\textbf{Current-biased $V(I)$ characteristic for a 6.7\,$\mu$m wide, 339\,nm thick BNCD microbridge at 1.9\,K.} Only the positive current branch is shown for clarity.
\textbf{(a)} $I_c$ and $I_R$ denote the critical current and retrapping current respectively. The arrows show the current direction around a hysteretic loop.
\textbf{(b)} An expanded view of the falling current branch showing discrete voltage steps with subsequent linear slopes that converge to the same point $I_{0}=13.78\,\mu A$ on the current axis. 
\textbf{(c)} A detailed view of individual voltage steps, showing the converging linear slopes.
} 
\end{figure} 

The observed signature of low-dimensional superconductivity in large BNCD microbridges was further examined by measuring the current-biased $V(I)$ characteristics at 1.9\,K, sufficiently far below $T_c$ that the residual resistance is negligible. The three microbridges display the same general characteristics and behavior so here we will focus on the 6.7\,$\mu$m wide bridge for which we have the most detailed measurements and analysis.
 
The $V(I)$ characteristic, the positive current branch of which is shown in Fig.~\ref{IVC}a, is strongly hysteretic and shows underdamped Josephson-like behavior. When the critical current, $I_c = 96.6\,\mu$A, is reached there is a large and discontinuous jump to a high-voltage state with a resistance close to the normal state resistance. Once in the normal state, as the current is reduced, a hysteretic return path is followed, and the superconducting state is re-established at the so-called retrapping current, $I_R = 21.4$\,$\mu$A. We note that, similar to the 1D Mo$_{79}$Ge$_{21}$ systems described by Sahu et al\cite{sahu2009individual}, $I_c$ appears to be stochastic, with a maximum observable value of $I_{c(max)} = 115$\,$\mu$A at 1.9\,K, while $I_R$ is completely reproducible. The stochastic nature of $I_c$ gives a range of critical current densities for the same microbridge.

On the reducing current return path, shown in greater detail in Fig.~\ref{IVC}b and Fig.~\ref{IVC}c, a large number of discrete voltage steps separated by linear slopes are clearly visible, the locations of which are highly reproducible. By fitting to these linear slopes, we find that they extrapolate back to a single point on the current axis, $I_{0}=13.78\,\mu$A. This behavior is strongly reminiscent of PSCs in one-dimensional superconductors. The existence of an intercept at $\sim 0.5-0.7\, I_R$ has its origin in the fact that PSCs involve a \emph{temporal} oscillation between states in which the current is alternately carried by supercurrent and normal current\cite{tinkham1979interaction,skocpol1974phase}. To our knowledge, such consistent behavior over a large number of voltage steps has not been reported in studies of Josephson junction arrays.

\subsection{Metastability in the current-voltage characteristics}

\begin{figure}
\includegraphics[width=\textwidth]{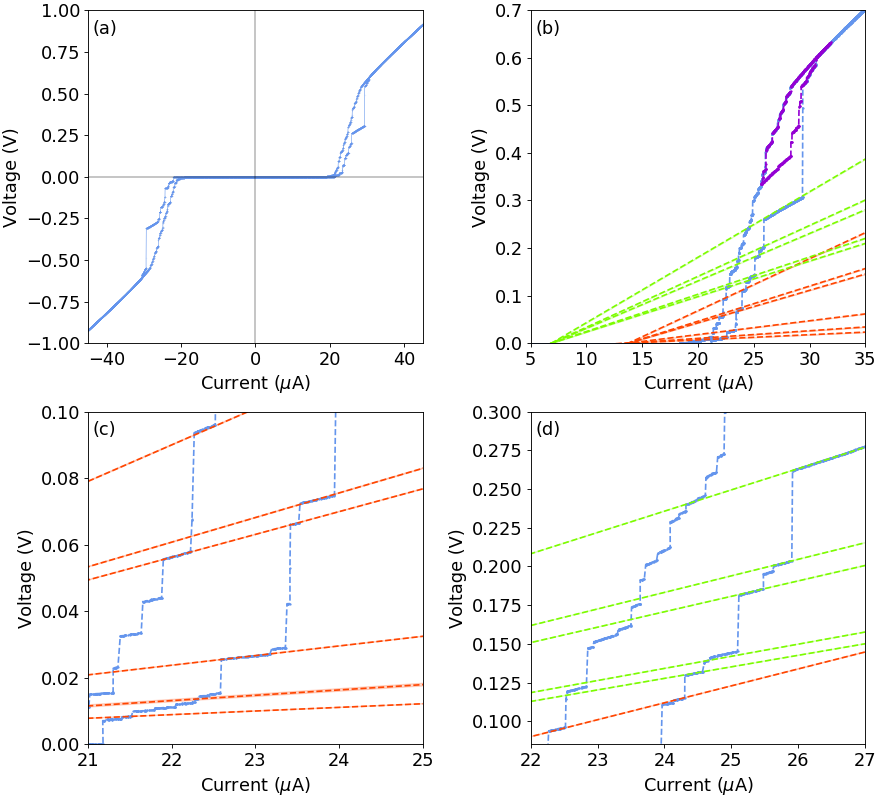}
\caption{
\label{Noise}
\textbf{$V(I)$ characteristic of the 6.7\,$\mu$m wide BNCD microbridge switched by electromagnetic noise.} 
\textbf{(a)} The full $V(I)$ characteristic measured at 1.9\,K for the \emph{same} sample as shown in Fig.~\ref{IVC} switched into the metastable resistive state near to $I_R$ by an applied electromagnetic noise pulse during the upward current sweep. 
\textbf{(b)} Detail of the hysteretic region close to $I_R$. A repeatable hysteretic loop (purple) can be traced if the bias current is not returned to zero after switching. \textbf{(c)}, \textbf{(d)} Details of individual current-voltage steps, showing converging linear slopes with selected fit lines for clarity. Red lines converge to $I_{0}=13.78\,\mu A$; green lines converge to $I_{1}=6.89\,\mu A$.}
\end{figure} 

In addition to the strongly hysteretic behavior shown in Fig.~\ref{IVC}a, we used an externally applied electromagnetic noise pulse to induce a completely new branch in the $V(I)$ characteristic. Fig.~\ref{Noise}a shows the full $V(I)$ characteristic for the \emph{same} 6.7\,$\mu$m wide microbridge shown in Fig.~\ref{IVC}, now under the influence of such a noise pulse initiated just above $I_R$ as the current is swept upward. The return current path is again highly reproducible, unchanged as compared to Fig.~\ref{IVC}, and is largely unaffected by the introduction of externally supplied electromagnetic noise, either continuous or pulsed. Upon \emph{noise switching}, rather than showing a single sharp upward jump at some $I_c \gg I_R$, the rising-current behavior now becomes qualitatively similar to the return path, with a much reduced hysteresis. We observe that the sample displays a metastability between the voltage-carrying state and the zero-voltage state and that it is possible to switch between the two. The behavior we observe is similar to that reported in superconducting $\delta$-MoN nanowires, which are clearly one-dimensional and where the resistance is dominated by PSCs\cite{buh2015control}, and in YBa$_2$Cu$_3$O$_{7-x}$ where it is suggested the resistance is dominated by PSLs\cite{lyatti2020energy}. This supports the suggestion that our samples are acting in a similar way to PSC/PSLs.

Fig.~\ref{Noise}b shows a detailed view of the {\it switched} branch of the $V(I)$ characteristic. There is still a pronounced hysteresis, but now the rising and falling current paths meet the current axis at approximately the same place. The minimum critical current, $I_{c(min)}=21.19\,\mu A$, is very close to $I_R$ and is approximately $0.18\,I_{c(max)}$ at 1.9\,K. At low bias currents close to $I_{c(min)}$, small discrete voltage steps and differential resistance slopes similar to those on the return path are observed, albeit with some differences in detail. While the return path is highly reproducible, the rising path shows small differences in the exact location of the voltage steps in the low current region on successive current ramps. Also, before reaching the normal state resistance at a bias current $I_n=30.6\,\mu A$, there is a reproducible longer linear slope that is qualitatively different to the preceding slopes and is observed in all three samples. 

The switched rising path is shown in more detail in Fig.~\ref{Noise}c and Fig.~\ref{Noise}d. As with the return path shown in Fig.~\ref{IVC}, there are discrete voltage steps separated by slopes of differential resistance which extrapolate to distinct non-zero points on the current axis, as is characteristic for PSCs\cite{li2015local,delacour2012quantum,tinkham1979interaction}. For voltages below approximately 100\,mV, the extrapolation of these slopes often pass through a corresponding step on the the return path, and intercept at the same point, $I_{0}=13.78\,\mu A$ (red lines). Above about 100\,mV, the slopes extrapolate to a new point, $I_{1}=6.89\,\mu A$, which appears to be $0.5\,I_{0}$ to a high precision (green lines). This includes the long linear slope which precedes a large voltage jump that takes the system almost into the normal resistive state. This extrapolation of differential resistance slopes to \emph{two} distinct points of origin on the current axis has not previously been observed.

\begin{figure}
\includegraphics[width=0.65\textwidth]{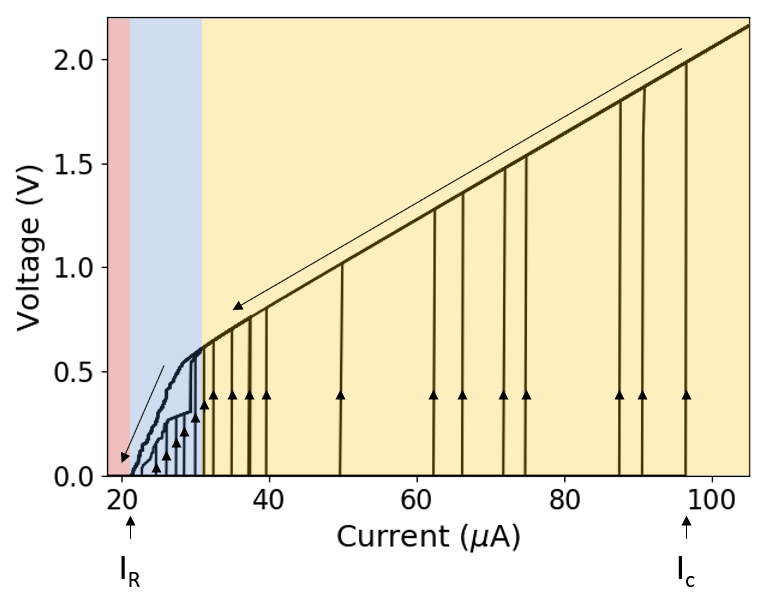}
\caption{\textbf{Controlled electromagnetic noise pulse induced switching of the 6.7\,$\mu$m wide BNCD microbridge state.}
\label{Switching}
The bias current through the microbridge at 1.9\,K is repeatedly ramped above $I_{c}$ and then back to below $I_R$. On each ramp the  noise pulse is applied to coincide with a different value of bias current on the rising path. The plot is a composite of all ramps; the points at which the pulses were applied are shown by the vertical arrows on the graph. The colored regions are as described in the main text.
} 
\end{figure}

To further explore the metastability of the voltage state, we performed repeated ramps of the bias current to above $I_{c(max)}$ and down again whilst applying a noise pulse at a range of points on the rising current part of each ramp. The results are shown in Fig.~\ref{Switching} as a composite of all bias current ramps. The values of bias current at which the pulses were applied are shown by the vertical arrows and the $V(I)$ characteristic has been divided into three regions of qualitatively different response. In the region $[0,I_R)$ (shaded red in Fig.~\ref{Switching}), the system is unaffected by noise pulses and is in the thermodynamically stable superconducting state. In the region $[I_n,I_{c(max)})$ (yellow), the zero voltage state is clearly metastable and is switched into the stable normal state by the application of a pulse. In the region $[I_R,I_n)$ (blue), the zero voltage state is again metastable and the pulses take the voltage up to the rising branch that can be initiated by a pulse just above $I_R$. Once the rising branch is accessed by a single pulse, the entire upward path is then followed with no need for any further pulses. This rising branch represents a non-normal voltage-carrying state which must be of a lower energy than the superconducting state for this current range. We have noted that, once initiated by a pulse close to $I_R$, the rising branch is slightly different on successive upward sweeps. The same is true for application of pulses anywhere in the blue region, i.e. pulses take the system up to slightly different voltages on successive ramps. There are clearly a number of metastable states in this region. The exception appears to be the long linear slope that precedes the large voltage jump into close to the normal state which, like the return path, is highly reproducible. 

There are other potential minima on the rising path which can only be accessed by sweeping the current above $I_n$, reducing it to a value in the range $(I_R,I_n)$, and then ramping it back up, whereupon it follows a previously inaccessible voltage-carrying path. The purple loop in Fig.~\ref{Noise}b is one example of such a hysteresis loop. The high degree of reproducibility of the return path compared to the rising path suggests that the return path possesses the deeper potential minimum. That the more stable branch is the one with the higher voltage would indicate that PSC/PSLs are energetically favorable in the region $(I_R,I_n)$, but cannot be formed on the rising current path without the introduction of additional energy in the form of noise, due to some free energy activation barrier, in agreement with the form of the $R(T)$ characteristic below $T_c$.

\section{Discussion and Conclusion}
\subsection{Origins of the phase slip behavior}

Having observed the characteristic features of phase slip physics in these BNCD microbridges, we must ask how these could be present and what form they take. Recall that both the physical dimensions of the system in comparison to the coherence length, and the shape of the $R(T)$ curve in the normal state confirm that these bridges are truly three-dimensional in nature. As PSCs are a 1D phenomenon and PSLs are 2D, some interpretation is required.

One striking feature that is explained by a phase-slip interpretation is that the slopes of the linear ramps between steps extrapolate to one of two common intercepts on the current axis. Within the phase slip interpretation, the extrapolation back to a common intercept is predicted by the theory of Skocpol, Beasley and Tinkham\cite{skocpol1974phase,tinkham1979interaction}, which considers the voltage across a PSC due to quasiparticle diffusion. In this model the voltage produced by $n$ PSCs is 

\begin{equation}
V(I)=R(I-I_0)\frac{2n\Lambda}{L}\tanh \left(\frac{L}{2n\Lambda}\right),
\label{SBTeqn}
\end{equation}
where $R$ is the normal state resistance of the track, $n$ is the number of PSCs, $L$ is the track length, $\Lambda$ is the quasiparticle diffusion length, and $I_0\sim I_c/2$ is the intercept of the slopes on the current axis - determined by the time-average supercurrent at the center of the PSC. Although this model was developed for PSCs in 1D wires, as it essentially describes a situation where the supercurrent is interrupted by localized quasi-normal region, it could equally well be applied to PSLs although the parameters will no longer have the same physical significance. In this model, the voltage and concomitant differential resistance both increase with increasing $n$ leading to a family of straight lines tracing back to $I_0$.

In our system, there is a single intercept, $I_0=13.78\,\mu$A, on the \textit{return} path, for which the differential resistance for small $n$ predicts $\Lambda\sim1.6\,\mu$m. The switched branch of the rising current path develops a second point of intercept $I_1=6.89\,\mu$m$=0.5\,I_0$. We have argued above that the rising path is not a potential minimum which suggests that some additional physical process is at work. The large plateau on the rising path seen between 25 and 30\,$\mu$A in Fig.~\ref{Noise}b is highly reminiscent of resonant steps seen in Josephson junction arrays (JJAs)\cite{trias1997intrinsic} which, as we shall discuss, suggests that this additional physics could be related to some underlying collective JJA behavior.

We have argued above that the microstructure of BNCD is structurally similar to a JJA. Underdamped JJAs can produce voltage steps themselves due to the phenomenon of synchronized row switching\cite{phillips1994dynamics,trias1997intrinsic,ochs1996spatially,van1993vortex}. This occurs when an entire row of junctions perpendicular to the flow of current switches into (or out of) the normal state. This scenario is supported by both a wealth of theoretical modeling\cite{yu1992resistance,yu1992fractional} and low temperature scanning laser microscopy measurements\cite{ochs1996spatially}. Coherence across the array is maintained, and the switched row behaves as a single junction with critical current equal to the number of junctions in a row multiplied by the critical current per junction. Row switched states are conceptually very similar to PSLs, being a thin band of suppressed order parameter stretching across the width of the system. It may therefore be instructive to think of a PSL as being akin to the continuum limit of a row switched state. Indeed, Lognevov et al\cite{logvenov1996two} have argued that a switched row acts like a phase slip line in wide superconducting films for sufficiently high temperatures such that the vortex mass vanishes.

Notwithstanding row switching, phase slip behavior has been observed by previous authors in both 1D chains\cite{pop2010measurement,manucharyan2012evidence} and 2D arrays\cite{van1988coherent} of Josephson junctions. We therefore conclude that the origin of the behavior seen in these experiments is an accumulation of macroscopic PSLs forming across the width of the bridge, facilitated by the intrinsic JJA making up the microstructure of BNCD. We note that the sheet resistance for our sample is only $\sim100\,\Omega$ and this sets the resistance scale for the intergranular conductance in our samples. This is is much less than the resistance quantum ($\sim6\,$k$\Omega$) and should put our samples well into the metallic state\cite{yu1992resistance}, supporting a PSL interpretation. The fact that our samples are disordered should not be a bar to this behavior. In either a JJA or PSL model, would expect breakdown to occur along the lowest energy track, which may, or may not be perpendicular to the track length.

We have demonstrated the ability to switch between metastable voltage-carrying states, a feature which has previously been reported in nanowires\cite{buh2015control,lyatti2020energy}. As in these previous reports, the ability to switch is very promising for a number of applications. These include implementation as a pulse-controlled memory device\cite{buh2015control}, and as a novel circuit element for quantum sensing and computing\cite{mooij2006superconducting,lyatti2020energy}. The BNCD system has the practical advantage that it is not necessary to fabricate individual tunnel junctions to achieve these ends. A further advantage is that, in comparison with its 1D counterparts, the voltage across the device is orders of magnitude greater, which leads to a clearer readout. There is ongoing interest in quantum phase slips as these are the most promising route to a quantum current standard\cite{wang2019towards}. Given the ease with which PSLs are induced in this system, future work will look to create equivalent \textit{quantum} PSLs in similar structures.

\section{Conclusion}

In summary, we have measured the transport properties of three-dimensional BNCD microbridges and have found clear hallmarks of phase slip phenomena, which can be induced by the application of electromagnetic noise pulses. Measurement of the $R(T)$ and $V(I)$ characteristics performed on the same sample both support the conclusion that these microbridges carry macroscopic excitations akin to phase slips. We suggest that the origin of these macroscopic phase slips is tied to an intrinsic Josephson junction array formed by the columnar growth of nanocrystalline diamond.\\


\noindent Supplementary Information (BNCD film preparation, microbridge fabrication, measurement techniques, and additional data showing results from all three microbridges) is available.

\section*{Acknowledgments}
G.M.K. would like to thank R. Tucker, A. Harrison, C. Pakes, and M. Bose for useful discussions. The authors gratefully acknowledge financial support by the European Research Council under the EU Consolidator Grant `SUPERNEMS' (project ID 647471).

\section*{Competing financial interests}
The authors declare no competing financial interests.



\providecommand{\noopsort}[1]{}\providecommand{\singleletter}[1]{#1}%

\clearpage

\begin{center}
{\Large Supplemental Materials: \\ Phase Slips and Metastability in Granular Boron-doped Nanocrystalline Diamond Microbridges}
\end{center}

\setcounter{figure}{0}
\renewcommand\thefigure{S\arabic{figure}}

\subsection*{BNCD film preparation}{\label{Sample_preparation}}

The boron-doped nanocrystalline diamond (BNCD) film was grown on a 2-inch SC-1 cleaned (100) silicon wafer with a 500\,nm thick SiO$_2$ buffer layer using microwave plasma-assisted chemical vapour deposition in a Seki AX6500 series microwave plasma reactor system\cite{williams2008growth}. Before loading into the reactor chamber, the substrate was seeded for 10\,min by ultrasonic agitation in a monodisperse aqueous colloid of nanodiamond particles\cite{williams2007enhanced}, rinsed in deionised water, and spun dry.  The film was grown in a gas mixture of 3$\%$ methane in hydrogen. The chamber pressure was 40\,Torr, the microwave power was 3.5\,kW, and the substrate temperature was 720$^{\degree}$C during growth as measured by a dual wavelength pyrometer. The charge carriers required for superconductivity in diamond were provided by the addition of trimethylboron to the gas mixture with a B/C ratio of 12,800\,ppm. If 100\% incorporation efficiency from the gas phase is assumed, there is an upper limit of $2.3\times10^{21}$\,cm$^{-3}$ for the carrier concentration. The film thickness was monitored during growth by \textit{in situ} laser interferometry and subsequently measured by cross sectional scanning electron microscopy (SEM). The surface of the resulting 339\,nm thick film is shown in Fig.~1. The film has a distribution of grain sizes, with a mean grain size of 102\,nm and a surface roughness of $\sim$25\,nm RMS. Previous work has shown that the surface roughness of the film does not affect the shape and magnitude of the superconducting transition\cite{klemencic2017superconductivity}.

\subsection*{Microbridge fabrication}

The 1600\,$\mu$m long microbridges of varying width (4.6, 6.7, and 10.5\,$\mu$m) were fabricated from the 339\,nm thick BNCD film using a standard top-down photolithographic approach. Microbridge structures were defined by thermal evaporation of a 70\,nm nickel mask through patterned resist and subsequent lift-off in acetone. The exposed surrounding BNCD was etched by oxygen inductively coupled plasma reactive ion etching (ICP-RIE). The ICP power was 1500\,W, the RIE power was 100\,W, and the oxygen flow was 40\,sccm. The metal mask was removed in FeCl$_3$ and thoroughly cleaned to remove all traces of the mask. The unpatterned sample is an approximately 5\,mm square piece of the BNCD sample with the current and voltage leads arranged in a Van der Pauw configuration. 

\subsection*{Measurement Techniques}{\label{Measurement Techniques}}

Four-wire electrical contact to the microbridge samples was made by direct wire bonding to the surface and Ohmic contact behavior was confirmed. Resistance as a function of temperature, shown in Fig.~2, was measured using a Quantum Design Physical Property Measurement System. The excitation current was 0.5\,$\mu$A, well below both $I_{c(min)}$ and $I_{c(max)}$ of all microbridges.

$V(I)$ characteristics were measured at 1.9\,K in zero applied field using a bespoke battery-operated voltage-controlled current source and amplifier. The general behavior shown in Fig.~4 persists down to dilution refrigerator temperatures though with a corresponding increase in the achievable $I_{c(max)}$. Electromagnetic noise pulses were applied by either operating a small electrical motor close to the electrical measurement apparatus, or by summing the input bias signal with a pulse output by a second computer-controlled arbitrary waveform generator.

\subsection*{Additional data}

\begin{figure}
\includegraphics[width=\textwidth]{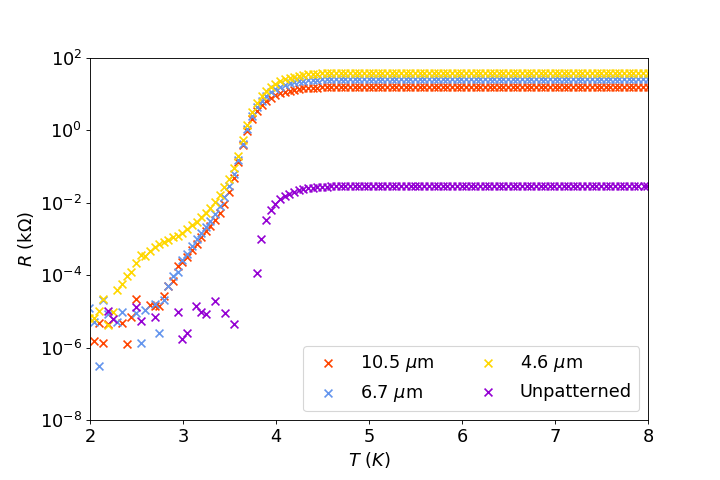}
\caption{Resistance as a function of temperature, $R(T)$, for BNCD microbridges and the unpatterned film shown for the range 2--8\,K.  }
\label{FigS1}
\end{figure}

\begin{figure}
\includegraphics[width=\textwidth]{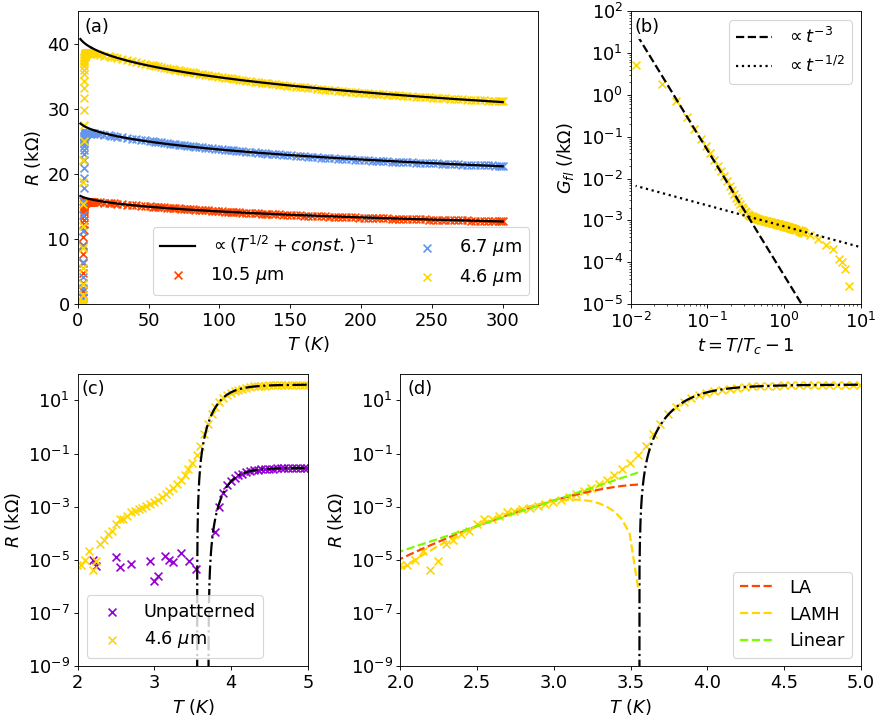}
\caption{\textbf{Resistance as a function of temperature, $R(T)$, for BNCD microbridges - focusing on the 4.6\,$\mu$m wide microbridge - and the unpatterned film.}
\textbf{(a)} $R(T)$ for three 1600\,$\mu$m long microbridges (widths as indicated by the legend) in the temperature region 2 -- 300\,K. 
The high temperature data are fit to the form $G(T)=a+b\sqrt{T}$ expected from electron-electron interaction theory. 
\textbf{(b)} Log-log plot of the fluctuation conductivity $G_{fl}(T)$ as a function of the reduced temperature, showing a crossover from 0D to 3D behavior at $T-T_c\sim 0.3$\,K.
\textbf{(c)} The superconducting transition of the unpatterned film does not show evidence of broadening below $T_c$. The black line is a fit to the fluctuation conductivity. The superconducting transition of the 4.6\,$\mu$m wide microbridge is shown for comparison.
\textbf{(d)} Low temperature $R(T)$ for the 4.6\,$\mu$m wide microbridge. The resistance below $T_c$ is fit to the LA (red), LAMH (yellow) and
linear (green) forms. Note that none of these are expected to be accurate close to $T_c$ as they assume large free energy barriers.}
\label{FigS2}
\end{figure}

\begin{figure}
\includegraphics[width=\textwidth]{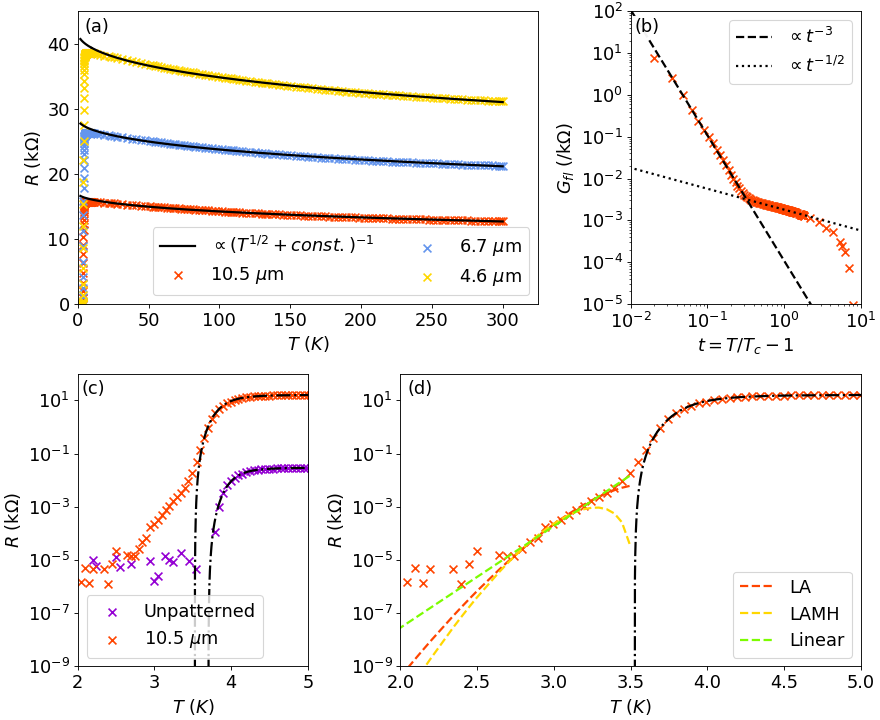}
\caption{\textbf{Resistance as a function of temperature, $R(T)$, for BNCD microbridges - focusing on the 10.5\,$\mu$m wide microbridge - and the unpatterned film.}
\textbf{(a)} $R(T)$ for three 1600\,$\mu$m long microbridges (widths as indicated by the legend) in the temperature region 2 -- 300\,K. 
The high temperature data are fit to the form $G(T)=a+b\sqrt{T}$ expected from electron-electron interaction theory. 
\textbf{(b)} Log-log plot of the fluctuation conductivity $G_{fl}(T)$ as a function of the reduced temperature, showing a crossover from 0D to 3D behavior at $T-T_c\sim 0.3$\,K.
\textbf{(c)} The superconducting transition of the unpatterned film does not show evidence of broadening below $T_c$. The black line is a fit to the fluctuation conductivity. The superconducting transition of the 10.5\,$\mu$m wide microbridge is shown for comparison.
\textbf{(d)} Low temperature $R(T)$ for the 10.5\,$\mu$m wide microbridge. The resistance below $T_c$ is fit to the LA (red), LAMH (yellow) and
linear (green) forms. Note that none of these are expected to be accurate close to $T_c$ as they assume large free energy barriers.}
\label{FigS3}
\end{figure}

\begin{figure}
\includegraphics[width=\textwidth]{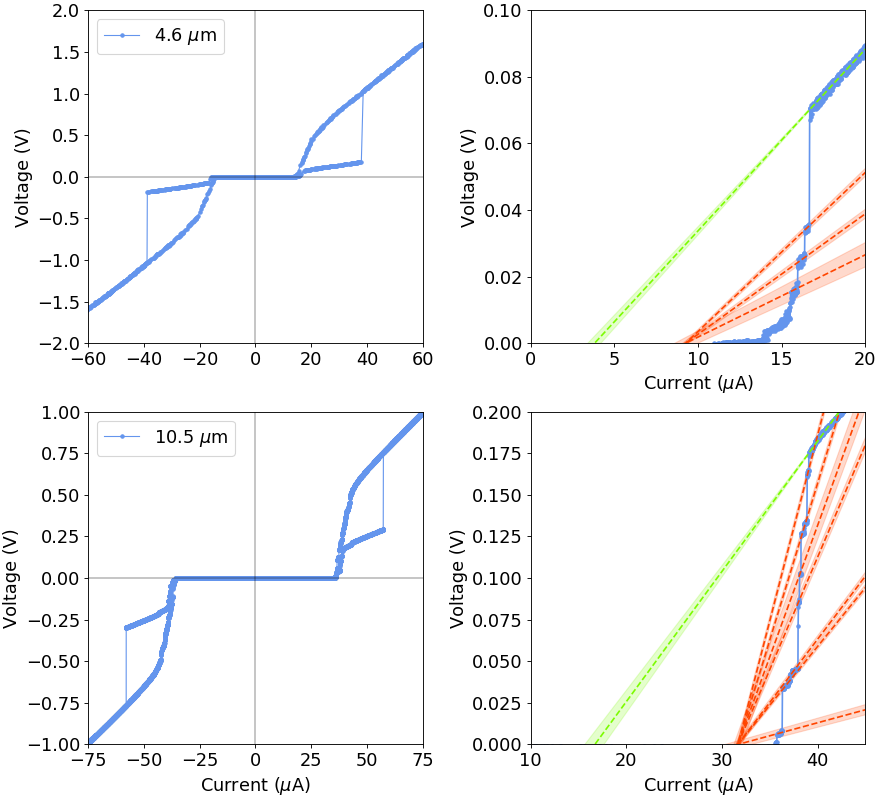}
\caption{\textbf{Current-biased $V(I)$ characteristics for the 4.6\,$\mu$m wide (top) and 10.5\,$\mu$m wide (bottom) microbridges.}
A detailed view of individual voltage steps for each sample, showing the converging linear slopes, is shown next to the full $V(I)$ characteristic for each sample. As with the 6.7\,$\mu$m wide microbridge, there are two points of intercept on the current axis, where the lower intercept is around half of the upper.}
\label{FigS4}
\end{figure}

For completeness, the $R(T)$ and $V(I)$ graphs for the 4.6 and 10.5\,$\mu$m microbridges are shown here. Fig.~\ref{FigS1} (with detailed fits shown in Fig.~\ref{FigS2} and \ref{FigS3}) shows all three microbridge $R(T)$ curves together with that of the unpatterned film; an additional broadening is observed below $T_c$ for the microbridge samples. This data does show a systematic dependence on bridge width, although it is impossible to make conclusive statements based on only three values of width. The phase slip shoulder in the $R(T)$ data becomes narrower as bridge width increases, as expected theoretically since the free energy barrier for a phase slip increases with width. Fig.~\ref{FigS4} shows $V(I)$ data for these same microbridges, showing that similar discrete voltage steps are formed. Similarly, the current intercept $I_0$ in the $V(I)$ data increases with bridge width, again because the free energy barrier increases with width.

\providecommand{\noopsort}[1]{}\providecommand{\singleletter}[1]{#1}%

\end{document}